\documentclass[a4paper,twocolumn,english]{revtex4-2}
\usepackage[T1]{fontenc}
\usepackage[latin9]{inputenc}
\usepackage{float}
\usepackage{amsmath}
\usepackage{graphicx}
\usepackage{esint}

\makeatletter

\usepackage[labelsep=period]{caption}
\renewcommand{\fnum@figure}{FIG.~\thefigure}

\@ifundefined{showcaptionsetup}{}{%
 \PassOptionsToPackage{caption=false}{subfig}}
\usepackage{subfig}
\makeatother

\usepackage{babel}
\begin{document}
\title{Heuristic approach to trajectory correlation functions in bounded
regions with Lambert scattering walls}
\author{T. Rao}
\author{R. Golub}
\affiliation{Physics Department, North Carolina State University, Raleigh, North
Carolina 27695, USA}
\begin{abstract}
The behavior of spins undergoing Lamor precession in the presence
of time varying fields is of interest to many research fields. The
frequency shifts and relaxation resulting from these fields are related
to their power spectrum and can be determined from the Fourier Transform
of the auto-correlation functions of the time varying field. Using
the method of images {[}C. M. Swank, A. K. Petukhov, and R. Golub,
Phys. Lett. A \textbf{376}, 2319 (2012){]} calculated the position-position
auto-correlation function for particles moving in a rectangular cell
with specular scattering walls. In this work we present a heuristic
model that extends this work to the case of Lambert scattering walls.
The results of this model are compared to simulation and show good
agreement from the ballistic to diffusive regime of gas collisions,
for both square and general rectangular cells. This model requires
three parameters, two of which describe the distribution of images
in the case of a square cell, and one of which describes the asymmetry
in the mixing of the x and y components of the velocity in the case
of non-square rectangular cells.
\end{abstract}
\maketitle

\section*{Introduction}

Systems of spins moving under the influence of static and time varying
magnetic fields are of wide-ranging scientific interest. It was Bloembergen,
Purcell, and Pound \citep{key-1} who first showed that the relaxation
time of the spins is given by the power spectrum of the fluctuating
field evaluated at the Lamor frequency. In general, the power spectra
can be written in terms of the Fourier transform of the auto-correlation
function of the fluctuating field. See the introductions of \citep{key-2,key-3}
for a brief history of the field.

One of the applications of these techniques is in next generation
neutron electric dipole moment (nEDM) searches, which require measurements
of spin dynamics at the nanohertz level. Additionally the motional
magnetic field resulting from the interaction between the spins and
the electric field in such experiments results in ``false EDM''
systematic error that needs to be accounted for \citep{key-7}. Redfield
\citep{key-4}, Slichter \citep{key-5}, and McGregor\citep{key-6}
have given formal derivations of the relation between relaxation and
the auto-correlation functions of the fields, and the method was applied
to the \textquotedblleft false EDM\textquotedblright \citep{key-7}
systematic error in nEDM searches \citep{key-8}. General methods
for calculating auto-correlation functions for particles diffusing
in inhomogenous fields can be found in \citep{key-9,key-10}.

Additionally, methods for calculating the trajectory auto-correlation
functions, or the frequency shift resulting from the motional magnetic
field have been developed for a variety of geometries, wall conditions,
and collision rates. Barabanov et al \citep{key-11} used a calculation
of the velocity-velocity correlation function for indiviual ballistic
trajectories to obtain the motional field frequency shift in the case
of a circular cell with specular scattering walls. This result was
expanded to include the effect of gas collisions by noting the spectrum
was a sum of hamonic oscillators, and including a damping term to
model the gas collisions. The work of Golub, Steyerl et al \citep{key-12,key-18}
similarly considered cylindrical cells in terms of individual ballistic
trajectories, but now with diffuse Lambert scattering walls. Additionally,
the case of square and more general rectangular cells for specular
and Lambert scattering walls were solved for. However, noticeable
qualitative differences between their calculation and simulation exist
in the case of non-square rectangular cells with Lambert scattering
walls. Specifically, simulation shows two peaks in the spectrum, while
their calculation only shows one. Clayton \citep{key-13} also considers
the case of reactangular cells, but in the diffusive limit rather
than the ballistic limit, and uses the condtional probabality of finding
the particle at a position inside the cell to calculate the trajectory
correlation functions. This work covers the 1D, 2D, and 3D cells but
only works when gas collisions are frequent enough that the diffusion
equation is valid. This work was expanded upon by Swank et al \citep{key-14,key-15}
giving solutions for 1D, 2D, and 3D rectangular cells that apply from
the ballistic to diffusive limit, but only in the case of specular
wall collisions. 

The method in \citep{key-14,key-15} utilizes the method of images
by applying the conditional probability of finding a particle at position
$r$, given initial position $r_{0}$, and travel time $\tau$, for
an unbounded domain \citep{key-16}. As explained in Fig. 1 going
from the real cell to the image cells can be represented by a triangle
wave periodic extension $\vec{\tilde{r}}_{0}(r_{0})$.

In this work we extend the technique of Swank et al to non specular
wall conditions, specifically the case of Lambert scattering walls.
We alter the position of the images in an attempt to alter the scattering
angle at the wall. This is accomplished by introducing a distribution
of reflecting wall positions. The properties of the image distribution
required to reproduce Lambert scattering is determined by comparing
the resulting spectrum of the trajectory correlation function to simulation.
In this way we develop an heuristic model of 2D rectangular cells
with Lambert scattering walls that's valid from the ballistic to diffusive
regime. Previously, Lambert scattering walls have only been considered
in the ballistic limit \citep{key-12}, 
\begin{figure}[H]
\includegraphics[width=8.6cm]{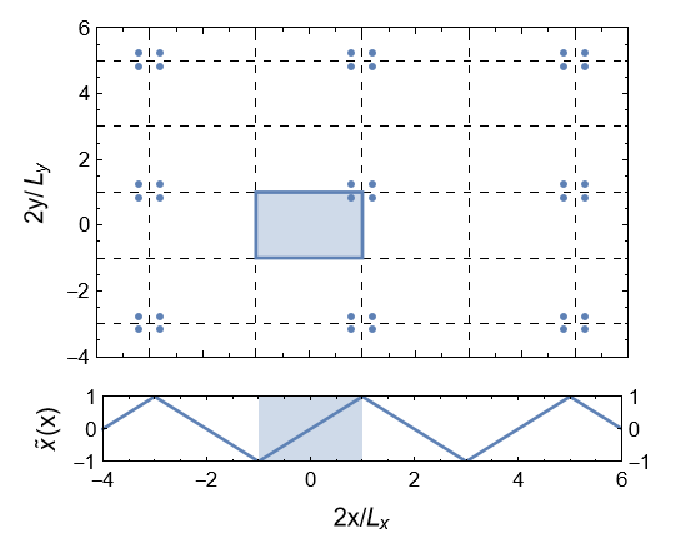}

\caption{The physical region (shaded) and a few of the periodic image cells
for the case of specular scattering walls described in \citep{key-14}.
The shown image positions all correspond to the same source inside
the physical cell. Trajectories between an image and an observation
point correspond to a trajectory between the source point and the
observation point that has undergone the same number of wall collisions
as image cell walls interesected by the image trajectory\emph{. }The
walls of the physical cell are located at $x=\pm L_{x}/2$ and $y=\pm L_{y}/2$.
The field in the physical region, in this case $B(x)=x$, is mirrored
at the boundary between image cells so that images see the same field
as the source. This results in a periodic extension of the position
inside the cell $\tilde{x}(x)$, in this case a traingle wave, over
the infinite domain.}
\end{figure}
giving results that agree with simulation in the case of cylindrical
and square cells, but not in the case of rectangular cells. 

\section*{The Position-Position Trajectory Correlation Function for Lambert
Scattering Walls}

From \citep{key-14} Eq. (10) we see we can write the position-position
trajectory correlation function as

{\small{}
\begin{equation}
\left\langle \overrightarrow{\widetilde{r}}_{0}(t)\cdot\overrightarrow{r}(t+\tau)\right\rangle =\frac{1}{L_{x}L_{y}}\int\int d^{2}r_{0}d^{2}r\overrightarrow{\widetilde{r}}_{0}(r_{0})\cdot\overrightarrow{r}P(\vec{r}-\vec{r_{0}},\tau)
\end{equation}
}with the infinite domain conditional probability $P(\vec{r}-\vec{r_{0}},\tau)$
from \citep{key-16}. The method of images allows the use of the infinite
domain conditional probability. The conditional probability of a trajectory
in a finite domain is replaced by the infinite domain conditional
probability between the image position $r_{0}$, and the observation
point inside the physical cell $r$, for travel time $\tau$. The
dot product is evaluated using the periodic extension of the physical
cell $\vec{\tilde{r}}_{0}(r_{0})$ for the source position. 

From Eq. (1) the 1D correlation function from \citep{key-14,key-15}
for a 2D cell can be written as

\begin{equation}
\left\langle \widetilde{x}_{0}(t)x(t+\tau)\right\rangle =\frac{1}{L_{x}}\intop_{-\infty}^{\infty}\intop_{-L_{x}/2}^{L_{x}/2}dx_{0}dx\widetilde{x}_{0}(x_{0})xP(\vec{r}-\vec{r_{0}},\tau)
\end{equation}

To get the Spectrum we must take the Fourier transform of Eq. (1).
From \citep{key-16} we know the Fourier-Laplace transform of the
2D infinite domain conditional probability is given by

\begin{equation}
P(\vec{q},s)=\frac{t_{c}}{\sqrt{(t_{c}vq)^{2}+(1+st_{c})^{2}}-1}
\end{equation}

where $t_{c}$ is the mean gas collisions time, and $v$ is the particle
speed. If we use an even extension of the conditional probability
to negative $\tau$ the Fourier transform of $P(\vec{r}-\vec{r_{0}},\tau)$
is a real even function equal to $2\mathrm{Re}[\intop_{0}^{\infty}P(r,t)e^{-i\omega t}dt]$.
This is simply twice the real part of the Laplace transform of $P(r,t)$
for $s=i\omega$ giving us

\begin{equation}
S_{xx}(\omega)=\frac{2}{L_{x}}\mathrm{Re}\left[\intop_{-\infty}^{\infty}\intop_{-L_{x}/2}^{L_{x}/2}dx_{0}dx\widetilde{x}_{0}(x_{0})xP(\vec{r}-\vec{r_{0}},i\omega)\right]
\end{equation}

where $S_{xx}(\omega)$ is the spectrum of 1D correlation function,
and $P(\vec{r}-\vec{r_{0}},s=i\omega)$ is the Laplace transform of
$P(\vec{r}-\vec{r_{0}},\tau)$. Additionally we can rewrite $P(\vec{r}-\vec{r_{0}},i\omega)$
as $\int d\vec{q}P(\vec{q},i\omega)\frac{e^{i\vec{q}\cdot(\vec{r}-\vec{r}_{0})}}{2\pi}$,
where $P(\vec{q},i\omega)$ is the Fourier-Laplace transform given
in Eq. (3). Since the conditional probability for an infinite domain
has no angular dependence we can choose $\vec{q}=q\hat{x}$ without
loss of generality. This in turn results in 

\[
S_{xx}(\omega)=\frac{2}{2\pi L_{x}}
\]

\begin{equation}
\mathrm{\times Re}\left[\intop_{-\infty}^{\infty}dq\intop_{-L_{x}/2}^{L_{x}/2}dx\intop_{-\infty}^{\infty}dx_{0}P(q,i\omega)e^{iq(x-x_{0})}\widetilde{x}_{0}x\right]
\end{equation}

or equivalently

\[
S_{xx}(\omega)=\frac{2}{2\pi L_{x}}
\]

\begin{equation}
\times\mathrm{Re}\left[\intop_{-\infty}^{\infty}dqP(q,i\omega)\intop_{-L_{x}/2}^{L_{x}/2}dxe^{iqx}x\intop_{-\infty}^{\infty}dx_{0}e^{iqx_{0}}\widetilde{x}_{0}\right]
\end{equation}

For specular walls, a trajectory's wall reflection is accounted for
by a single image position, since the reflected angle is exactly determined
by the incoming angle. For rough walls the reflected angle is random
and follows a cosine distribution. Because of this a single image
is no longer sufficient to describe the scattering. Instead there
exists a distribution of image positions that correspond to the possible
reflected trajectories (different observation points), or equivalently,
each image position can be thought of as corresponding to multiple
possible sources inside the physical cell. If the specular image is
altered by changing the size of the image cells, the $\widetilde{x}_{0}$
corresponding to a given image is also altered. We attempt to find
a distribution of image cell sizes that generate the correct distribution
of $\widetilde{x}_{0}$ to reproduce Lambert wall scattering when
calculating the trajectory correlation function. The distribution
of image cell sizes that corresponds to Lambert scattering can then
be used to calculate an effective extension to the x axis that can
be used in place of the specular periodic extension.\emph{ }

In \citep{key-14,key-15} the image positions for specular walls are
accounted for by a triangle wave extension of the particle's position
outside of the cell (See Fig. 1)

\begin{equation}
\widetilde{x}_{0}(x_{0})=\sum_{n=odd}\frac{-i^{n}2L_{x}}{\pi^{2}n^{2}}e^{i\frac{\pi n}{L_{x}}x_{0}}
\end{equation}

So for Lambert scattering walls we need an appropriate ensemble average
of triangle wave extensions that will reproduce the wall scattering.
This amounts to taking an ensemble average of triangle waves with
different periods $l$, which would correspond to the length of a
side of the physical cell in the case of specular scattering (Eq.
(8)). Additionally we average the x and y dimensions separately since
we want to include all combinations of $l_{y}$ and $l_{x}$. Note
that for a square cell the symmetry in the real cell suggests the
distribution of $l_{x}$ and $l_{y}$ should be the same. Consequently,
in the case of a square cell, although the individual image cells
being averaged over are rectangles, $l_{x}$ and $l_{y}$ are statistically
independent, the effective cell will still be a square.

\begin{equation}
\widetilde{x}_{0}(x_{0})=\sum_{odd}\left\langle \frac{-i^{n}2l}{\pi^{2}n^{2}}e^{i\frac{\pi n}{l}x_{0}}\right\rangle 
\end{equation}

If we assume a Gaussian distribution in $l$ with average value $L_{modx}$,
and with width $\sigma$, Eq. (8) becomes

\begin{equation}
\widetilde{x}_{0}(x_{0})=\sum_{odd}\frac{-i^{n}2}{\pi^{2}n^{2}}\intop_{0}^{\infty}le^{i\frac{n\pi}{l}x_{0}}\frac{e^{-\frac{(l-L_{modx})^{2}}{2\sigma^{2}}}}{\sqrt{2\pi}\sigma}dl
\end{equation}

Two examples comparing the x axis extension for Lambert scattering
walls given by Eq. (9) to the specular extension given by Eq. (7)
are plotted in Fig. 2.

\begin{figure}[H]
\subfloat[]{\includegraphics[width=8.6cm]{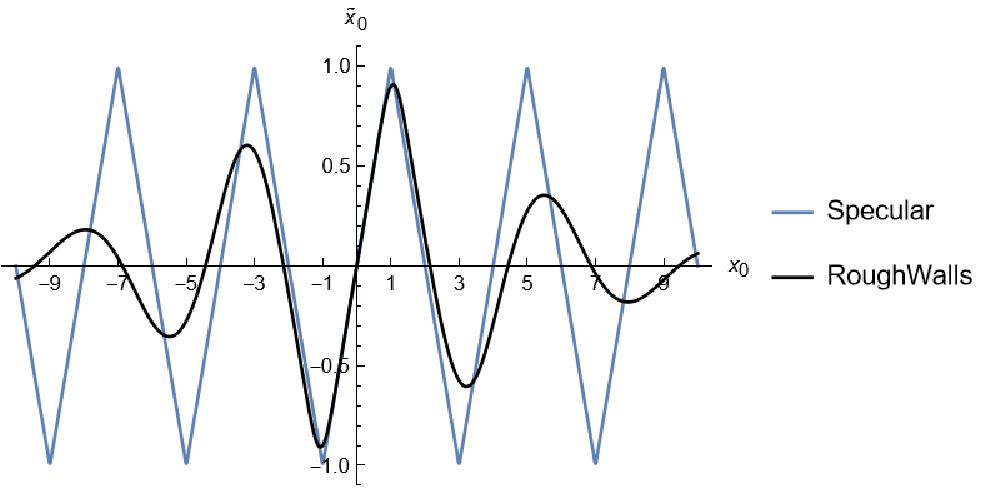}}

\subfloat[]{\includegraphics[width=8.6cm]{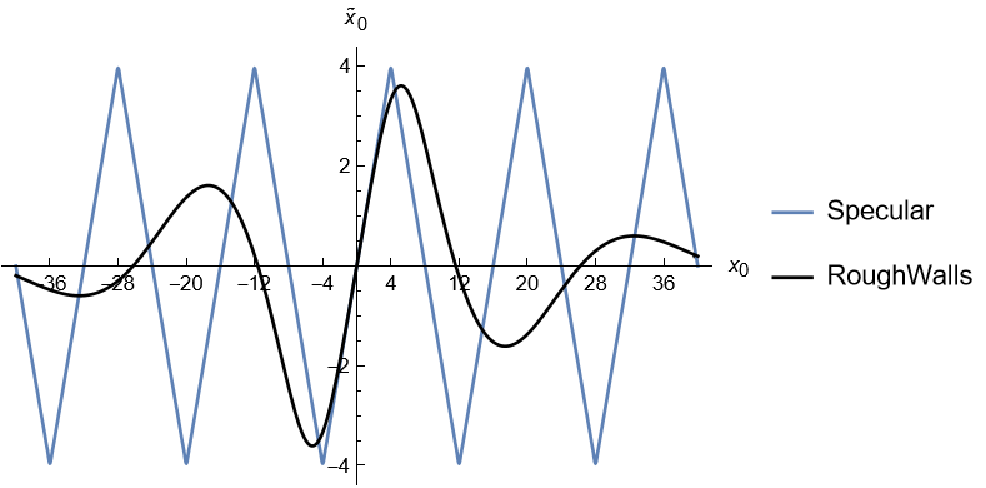}

}

\caption{Specular and rough wall extensions used to account for wall scattering
for a square and rectangular cell. (a) The specular and rough wall
extensions for a square cell with $L_{x}=2$, $L_{modx}=2.13$, and
$\sigma=0.393$. Note the region between -1 and 1 corresponding to
the physical cell is barely altered from the specular result. (b)
The specular and rough wall extensions for the long side of a rectangular
cell with an aspect ratio of 4, $L_{modx}=10.08$, and $\sigma=3.93$.
The length of the sides are normalized to 1/2 of the length of the
shorter side. Note the region between -4 and 4 corresponding to the
physical cell begins to significantly deviate from the specular case
near the walls of the physical cell.}
\end{figure}

In order to calculate the spectrum of the correlation function, Eq.
(6), we need to take the Fourier transform of Eq. (9) giving

\begin{equation}
\widetilde{x}_{0}(q)=\sum_{odd}\frac{-i^{n}2}{\pi^{2}n^{2}}\intop_{-\infty}^{\infty}\intop_{0}^{\infty}l\frac{e^{-\frac{(l-L_{modx})^{2}}{2\sigma^{2}}}}{\sqrt{2\pi}\sigma}e^{i\frac{n\pi}{l}x_{0}}e^{-iqx_{0}}dx_{0}dl
\end{equation}

evaluating the $x_{0}$ integral gives us

\begin{equation}
\widetilde{x}_{0}(q)=\sum_{odd}\frac{-i^{n}2}{\pi^{2}n^{2}}\intop_{0}^{\infty}l\frac{e^{-\frac{(l-L_{modx})^{2}}{2\sigma^{2}}}}{\sqrt{2\pi}\sigma}2\pi\delta(\frac{n\pi}{l}-q)dl
\end{equation}

we then use the substitution $u=\frac{n\pi}{l}$ and evaluate the
delta function

\begin{multline}
\widetilde{x}_{0}(q)=\sum_{odd}-i^{n}\frac{2\sqrt{2\pi}}{\sigma}\intop_{0}^{\infty}e^{-\frac{(\frac{n\pi}{u}-L_{modx})^{2}}{2\sigma^{2}}}\delta(u-q)\frac{1}{u^{3}}du\\
=\sum_{odd}-i^{n}\frac{2\sqrt{2\pi}}{\sigma\left|q\right|^{3}}e^{-\frac{(\frac{n\pi}{q}-L_{modx})^{2}}{2\sigma^{2}}}
\end{multline}

The magnitude of $q$ used in the denominator of Eq. (12) is needed
to account for the case of negative $n$ and consequently negative
$q$. We now define $q_{x0}=n\pi/L_{modx}$ and rewrite this as

\begin{equation}
\widetilde{x}_{0}(q)=\sum_{odd}-i^{n}\frac{4\pi}{L_{modx}q^{2}}\frac{n\pi}{\sqrt{2\pi}}\frac{e^{-\frac{n^{2}\pi^{2}(q_{_{x0}}-q)^{2}}{2(\sigma^{2}q^{2}q_{x0}^{2})}}}{(\left|q\right|\sigma q_{x0})}
\end{equation}

From Eq. (13) we now have $\int dxe^{-iqx_{0}}\widetilde{x}_{0}(x_{0})$

Additionally we can evaluate the x integral in Eq. (6) to get
\begin{equation}
\intop_{-L_{x}/2}^{L_{x}/2}dxe^{iqx}x=-\frac{i}{q^{2}}[qL_{x}\cos(\frac{qL_{x}}{2})-2\sin(\frac{qL_{x}}{2})]
\end{equation}

and we can now, using Eq. (13) and Eq. (14), write $S_{xx}$ , Eq.
(6), as

\begin{widetext}

\begin{equation}
S_{xx}=\frac{4}{L_{x}L_{modx}}\mathrm{Re}\left[\sum_{odd}i^{n+1}\intop_{-\infty}^{\infty}dqP(q,i\omega)\frac{1}{q^{4}}[qL_{x}\cos(\frac{qL_{x}}{2})-2\sin(\frac{qL_{x}}{2})]\frac{n\pi}{\sqrt{2\pi}}\frac{e^{-\frac{n^{2}\pi^{2}(q_{_{x0}}-q)^{2}}{2(\sigma^{2}q^{2}q_{x0}^{2})}}}{(\left|q\right|\sigma q_{x0})}\right]
\end{equation}

For a square $S_{xx}=S_{yy}$ and therefore the 2D spectrum $S_{rr}$
is simply $2S_{xx}$. For a more general rectangle we could write

\[
S_{rr}=\frac{4}{L_{x}L_{modx}}\mathrm{Re}\left[\sum_{odd}i^{n+1}\intop_{-\infty}^{\infty}dqP(q,i\omega)\frac{1}{q^{4}}[qL_{x}\cos(\frac{qL_{x}}{2})-2\sin(\frac{qL_{x}}{2})]\frac{n\pi}{\sqrt{2\pi}}\frac{e^{-\frac{n^{2}\pi^{2}(q_{_{x0}}-q)^{2}}{2(\sigma_{x}^{2}q^{2}q_{x0}^{2})}}}{(\left|q\right|\sigma_{x}q_{x0})}\right]
\]

\begin{equation}
+\frac{4}{L_{y}L_{mody}}\mathrm{Re}\left[\sum_{odd}i^{n+1}\intop_{-\infty}^{\infty}dqP(q,i\omega)\frac{1}{q^{4}}[qL_{y}\cos(\frac{qL_{y}}{2})-2\sin(\frac{qL_{y}}{2})]\frac{n\pi}{\sqrt{2\pi}}\frac{e^{-\frac{n^{2}\pi^{2}(q_{_{y0}}-q)^{2}}{2(\sigma_{y}^{2}q^{2}q_{y0}^{2})}}}{(\left|q\right|\sigma_{y}q_{y0})}\right]
\end{equation}

\end{widetext}

similarly to the case of specular walls \citep{key-14}. However,
unlike specular wall collisions which preserve the magnitude of $v_{x}$
and $v_{y}$ individually, Lambert wall collisions mix the x and y
components only preserving the overall magnitude of the velocity,
in the case of elastic collisions considered here. For a square cell
the symmetry between x and y means on average the x and y components
aren't effected by the velocity mixing. However, for rectangular cells,
$L_{y}>L_{x}$ and the velocity mixing is no longer symmetric. Each
wall collision results in change in the y velocity relative to the
x velocity which needs to be accounted for. Since the solution of
a square cell is given by $S_{xx}$ we can account for this effect
by altering the velocity used in calculating $S_{yy}$ relative to
the nominal velocity in $S_{xx}$. We note the velocity only appears
in the conditional probability and assume a Gaussian in effective
speed for calculating $S_{yy}$ giving

\begin{equation}
P(q,i\omega)\rightarrow\intop_{0}^{\infty}P(v,q,i\omega)\frac{e^{-\frac{(v-v_{nom})^{2}}{2\sigma_{v}^{2}}}}{\sqrt{2\pi}\sigma_{v}}dv
\end{equation}
Where the average speed $v_{nom}$ is determined from the ballistic
time $t_{b}$ for a square cell with sides of length $L_{x}$. The
ballistic time is taken to be the minimum time required to travel
from the center of a square cell to one of the walls. This then gives
$v_{nom}=\frac{L_{x}}{2t_{b}}$. For simplicity we will take $L_{x}=2$,
and measure times in units of $t_{b}$ so that $v_{nom}=1$.

However a Gaussian distribution cannot be exactly correct as the speed
cannot be less than zero, and a Gaussian allows for negative values
if the standard deviation is large enough. We will show that the width
of the Gaussian is $\approx0.4$ in the ballistic limit for a 2x8
cell. Which results in $\approx0.6$ \% of the Gaussian's area being
less than zero. By changing the integral so the distribution is cutoff
for $v<0.1v_{nom}$ we have now have 
\begin{equation}
\intop_{0.1}^{\infty}P(v,q,i\omega)\frac{e^{-\frac{(v-1)^{2}}{2\sigma_{v}^{2}}}}{\sqrt{2\pi}\sigma}dv=P_{v}(q,i\omega,\sigma_{v})
\end{equation}
 and problems at small $v$ due to the the unphysical Gaussian tail
are removed while only effecting $\approx1.2$ \% of the area of the
probability distribution for a 2x8 cell. Above $v$=0.1 the Gaussian
approximation works well.

Applying this change to the conditional probability we then have

\begin{widetext}

\[
S_{rr}=\frac{4}{L_{x}L_{modx}}\mathrm{Re}\left[\sum_{odd}i^{n+1}\intop_{-\infty}^{\infty}dqP(q,i\omega)\frac{1}{q^{4}}[qL_{x}\cos(\frac{qL_{x}}{2})-2\sin(\frac{qL_{x}}{2})]\frac{n\pi}{\sqrt{2\pi}}\frac{e^{-\frac{n^{2}\pi^{2}(q_{_{x0}}-q)^{2}}{2(\sigma_{x}^{2}q^{2}q_{x0}^{2})}}}{(\left|q\right|\sigma_{x}q_{x0})}\right]
\]

\begin{equation}
+\frac{4}{L_{y}L_{mody}}\mathrm{Re}\left[\sum_{odd}i^{n+1}\intop_{-\infty}^{\infty}dqP_{v}(q,i\omega,\sigma_{v})\frac{1}{q^{4}}[qL_{y}\cos(\frac{qL_{y}}{2})-2\sin(\frac{qL_{y}}{2})]\frac{n\pi}{\sqrt{2\pi}}\frac{e^{-\frac{n^{2}\pi^{2}(q_{_{y0}}-q)^{2}}{2(\sigma_{y}^{2}q^{2}q_{y0}^{2})}}}{(\left|q\right|\sigma_{y}q_{y0})}\right]
\end{equation}

\end{widetext}

Additionally we can recover the specular limit from Eq. (15) by taking
$\sigma_{x}$ goes to zero and $L_{modx}$ goes to $L_{x}$. Taking
this limit $\frac{n\pi}{\sqrt{2\pi}}\frac{e^{-\frac{n^{2}\pi^{2}(q_{_{0}}-q)^{2}}{2(\sigma^{2}q^{2}q_{0}^{2})}}}{(\left|q\right|\sigma q_{0})}\rightarrow\delta(q_{x0}-q)$,
giving 
\[
S_{xx}=-\frac{8}{L_{x}^{2}}\sum_{odd}\mathrm{Re}\left[P(q_{x0},i\omega)\left(\frac{1}{q_{x0}}\right)^{4}i^{n+1}\sin(n\pi/2)\right]
\]

\[
=\frac{8}{L_{x}^{2}}\sum_{odd}\mathrm{Re}\left[P(q_{x0},i\omega)\right]\left(\frac{1}{q_{x0}}\right)^{4},
\]
the expected specular result. In the limit of large gas collision
rate ($\lambda=t_{b}/t_{c}=1/t_{c}$) the wall collisions will be
suppressed by the gas collisions, and the Lambert scattering walls
and specular walls results should be equal. Therefore $\sigma$ in
this limit should tend toward zero and $L_{modx}$ to $L_{x}$. Additionally
for rectangular cells in this limit we should see that $\sigma_{v}$
goes to zero (as the velocity mixing from wall collisions becomes
negligible) so that $P_{v}\rightarrow P$.

\section*{Model of Gas Collision Rate Dependence of the Parameters of the Gaussians}

One way to consider the gas collision rate dependence of the correlation
function is that a trajectory that undergoes a gas collision but not
a wall collision should be accurately represented by the specular
solution, as it doesn't interact with the walls. Therefore as the
rate of gas collisions increase we should expect to see more trajectories
like this and the ensemble average of triangle wave extensions from
Eq. (8) will move toward a narrower distribution closer to being centered
on $L$. Therefore it's reasonable to think $L_{mod}$ and $\sigma$
should go like the fraction of wall collisions to total collisions,
which can be written as
\begin{equation}
N_{w}/N=\frac{N_{w}}{N_{w}+N_{g}}=\frac{1}{1+N_{g}/N_{w}}
\end{equation}

where $N_{w}$ is the number of wall collisions, N is the total number
of collisions, and $N_{g}$ is the number of gas collisions. In order
to understand the gas collisions' contribution let us consider the
number of gas collisions per wall collisions in a square cell with
sides $L$. Describing the gas collisions as a 2D random walk the
number of gas collisions that occur while traveling a distance $d$
is $N_{g}=(d/l_{mfp})^{2}$, where $l_{mfp}$ is the mean free path.
Given that $v=1$, and the gas collision rate $\lambda=1/t_{c}$,
then $l_{mfp}=vt_{c}=1/\lambda$. On average, a wall collision will
occur after traveling a distance $L/2$ in 1D. Therefore adjusting
for the the 4 walls (and therefore 4 wall collisions) of a square
cell, the number of gas collisions per wall collision in the 1D correlation
function for a 2D cell ($S_{xx}$ or $S_{yy}$) is

\begin{equation}
\frac{N_{w}}{N}=\frac{1}{1+\frac{1}{4}\left(\frac{L\lambda}{2}\right)^{2}}
\end{equation}

where L is the length of cell in the specified direction

We can now calculate the expected $\sigma$ for different aspect ratios
as a function of $\lambda$ if we take the value of $\sigma$ in the
ballistic limit and scale it by $N_{w}/N$ giving
\begin{equation}
\sigma(\lambda)=\sigma_{0}\frac{N_{w}}{N}
\end{equation}

Where $\sigma_{0}$ is the ballistic limit value of $\sigma$. Similarly,
the expected $L_{mod}(\lambda)$ would be

\begin{equation}
L_{mod}(\lambda)=(L_{mod0}-L)\frac{N_{w}}{N}+L=\text{\ensuremath{\Delta L_{0}}}\frac{N_{w}}{N}+L
\end{equation}

where $L_{mod0}$ is the ballistic limit value of $L_{mod}$, and
$\Delta L_{0}=L_{mod0}-L$. 

However, when extending this model of the $\lambda$ dependence to
work with $\sigma_{v}$ we need to remember that the width of the
$v$ Gaussian comes from asymmetry in the mixing of the x and y components
of the velocity, so that the Gaussian gets narrower not simply as
wall collisions are reduced but with an additional factor of the relative
number of gas collisions between the calculation of $S_{yy}$ (a square
with sides $L_{y}$) and $S_{xx}$ (a square with sides $L_{x})$
for a given $\lambda$ giving 
\begin{equation}
\sigma_{v}(\lambda)=\sigma_{v0}\frac{1}{1+\frac{1}{4}a^{2}\left(\frac{aL_{x}\lambda}{2}\right)^{2}}
\end{equation}
Where $a=\frac{L_{y}}{L_{x}}$, and $\sigma_{v0}$ is the ballistic
limit value of $\sigma_{v}$.

\section*{Model of Aspect Ratio Dependence in the Ballistic Limit}

Although the previous model provides a way to find the spectrum of
the correlation function as a function of $\lambda$ given the ballistic
limit correlation function; $\sigma_{0}$, $L_{mod0}$, and $\sigma_{v0}$
still need to be determined for each value of the aspect ratio ($a$).
By understanding how $\sigma_{0}$ , $L_{mod0}$, and $\sigma_{v0}$
scale with the aspect ratio of the physical cell it is possible to
write the spectrum of the correlation function in a way that depends
only on three parameters ($\sigma_{square}$, $L_{modsquare}$, and
$\sigma_{v,a=\infty}$).

\subsection*{$L_{mod0}$ scaling in terms of the square cell}

For a square cell, scaling the cell to a larger square should preserve
the change in $L_{x}$ $\Delta L_{x}$ as a fraction of $L_{x}$ in
order to keep the same relative change in image positions from the
specular case. In the case of a rectangular cell where $L_{y}>L_{x}$,
with aspect ratio integer $a=\frac{L_{y}}{L_{x}}$, we would need
$a$ $L_{x}\times L_{x}$ square cells along the y direction, as shown
in Fig. 3, to make our rectangle. Therefore we might expect to see
$\Delta L_{y}/L_{y}=a\Delta L_{x}/L_{x}$ for this more general case.
Assuming this is true we then have

\begin{multline}
L_{mody0}=\frac{\Delta L_{y0}}{L_{y}}L_{y}+L_{y}=a\frac{\Delta L_{x0}}{L_{x}}L_{y}+L_{y}\\
=a^{2}\Delta L_{x0}+aL_{x}
\end{multline}
for a rectangular cell in the ballistic limit, where $\Delta L_{x0}=L_{modsquare}-L$,
and $L_{modsquare}$ is the average image cell wall length for a square
cell with sides $L_{x}$ in the ballistic limit. 

\begin{figure}[H]
\includegraphics[height=8.6cm]{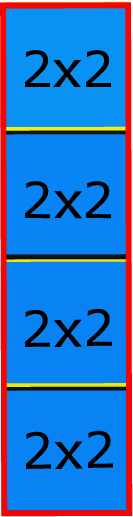}

\caption{Diagram of a single 2x8 periodic cell constructed from 4 2x2 image
cells. The ``real'' boundary of the cell is red. The boundary of
each 2x2 cell alternates between black or yellow to show the overlap
of the boundary between neighboring cells. Note that 6 of the 4 2x2
squares' walls used to make the rectangular cell are internal to the
rectangular cell.}
\end{figure}

From this same idea we would then also have that $\Delta L_{x}/L_{x}$
is independent of the aspect ratio. Also note that in these arguments
$a$ is always a positive integer as we are constructing the rectangular
cell in units of the $L_{x}\times L_{x}$cell. 

\subsection*{$\sigma_{y0}$ scaling in terms of the square cell}

For Lambert scattering walls each wall collision mixes the x and
y components of a trajectory. Therefore wall collisions along the
x (or y) axis also contribute to $\sigma_{y}$ (or $\sigma_{x}$).
So in determining $\sigma_{y0}$, we start by considering square space
$L_{y}\times L_{y}$ in size made from $L_{x}\times L_{x}$ image
cells in an attempt to include the collisions along the x axis that
contribute to $\sigma_{y0}$. Doing so we would then have

\begin{equation}
\sigma_{y0}=a^{2}\sigma_{square}
\end{equation}

Where $\sigma_{square}$ is the standard deviation of the image cell
wall length for the 2x2 square cell in the ballistic limit. However,
this expression fails to take into account the fact that scaling the
standard deviation as if we had $a^{2}$ $L_{x}\times L_{x}$ square
cells results in effectively over estimating the contribution of wall
collisions in the y direction. This scaling adds $2(a-1)$ unphysical
internal walls to each physical rectangular cell in the space. See
Fig. 3. If we assume every wall of a $L_{x}\times L_{x}$ cell equally
contributes $1/4\sigma_{square}$ to $\sigma_{y0}$, then since there
are $a$ rectangular physical cells over the $L_{y}\times L_{y}$
space, we find we have overestimated $\sigma_{y}$ by $1/2a(a-1)\sigma_{square}$.
Therefore the correct $\sigma_{y}$ is 

\begin{multline}
\sigma_{y0}=a^{2}\sigma_{square}-\frac{1}{2}(a^{2}-a)\sigma_{square}\\
=\frac{\sigma_{square}}{2}(a^{2}+a)
\end{multline}

\subsection*{$\sigma_{x0}$ scaling in terms of the square cell}

Similarly, simply taking $\sigma_{x0}=\sigma_{square}$, will also
overestimate the contribution to $\sigma_{x0}$ made by wall collisions
in the y direction. For $\sigma_{y0}$ we saw the over estimation
was $\frac{1}{2}(a^{2}-a)\sigma_{square}$ over a $L_{y}\times L_{y}$
square space. If we rescale this result for a $L_{x}\times L_{x}$
space, we find the overestimate is $\frac{1}{2}\frac{(a^{2}-a)}{a^{2}}\sigma_{square}$.
$\sigma_{x0}$is then given by

\begin{multline}
\sigma_{x0}=\sigma_{square}\left(1-\frac{\frac{1}{2}(a^{2}-a)}{a^{2}}\right)\\
=\frac{\sigma_{square}}{2}\frac{(a^{2}+a)}{a^{2}}
\end{multline}

\subsection*{$\sigma_{v0}$ scaling as a measure of the asymmetry of the cell}

As discussed earlier $\sigma_{v0}$ comes from the asymmetry of the
cell. In the case where $L_{y}\gg L_{x}$ the situation is effectively
1D and only the y component of the velocity contributes to transport
of particles across the cell, but with a spread in the effective velocity
in the y direction of the particles determined by the number of collisions
in the x direction. Conversely, in the square cell x and y must be
weighted equally so the effective velocity in the y direction is always
its nominal value of 1. Given the value at $a=\infty$, $\sigma_{v0}$
could therefore reasonably be scaled by a measure of the asymmetry
of the cell giving

\begin{multline}
\sigma_{v0}(a)=\sigma_{v,a=\infty}\frac{A_{yy}-A_{xy}}{A_{yy}}=\sigma_{v,a=\infty}\frac{a^{2}L_{x}^{2}-aL_{x}^{2}}{a^{2}L_{x}^{2}}\\
=\sigma_{v,a=\infty}\frac{a^{2}-a}{a^{2}}
\end{multline}

where $A_{yy}$ is the area of a $L_{y}\times L_{y}$ square and $A_{xy}$
is the area of a $L_{x}\times L_{y}$ rectangle.

\section*{Comparison to simulation}

Now that we have a model of the $\lambda$ and aspect ratio dependence,
we combine the Eqs. (22)-(24) with Eqs. (25), (27)-(29) to get 

\begin{equation}
\sigma_{y}=\sigma_{square}\frac{a^{2}+a}{2\left(1+\frac{1}{4}\left(\frac{aL_{x}\lambda}{2}\right)^{2}\right)}
\end{equation}

\begin{equation}
\sigma_{x}=\sigma_{square}\frac{a^{2}+a}{2a^{2}\left(1+\frac{1}{4}\left(\frac{L_{x}\lambda}{2}\right)^{2}\right)}
\end{equation}

\begin{equation}
L_{mody}=a^{2}\frac{L_{modsquare}-L_{x}}{1+\frac{1}{4}\left(\frac{aL_{x}\lambda}{2}\right)^{2}}+aL_{x}
\end{equation}

\begin{equation}
L_{modx}=\frac{L_{modsquare}-L_{x}}{1+\frac{1}{4}\left(\frac{L_{x}\lambda}{2}\right)^{2}}+L_{x}
\end{equation}

\begin{equation}
\sigma_{v}=\sigma_{v,a=\infty}\frac{a^{2}-a}{a^{2}}\frac{1}{1+\frac{1}{4}a^{2}\left(\frac{aL_{x}\lambda}{2}\right)^{2}}
\end{equation}

Leaving us with 3 free parameters $\sigma_{square}$, $L_{modsquare}$,
and $\sigma_{v,a=\infty}$. These parameters are determined by fitting
to simulation. 

The simulation calculates trajectories for a uniform distribution
of particles in the cell. Over each timestep trajectories have constant
velocity, with checks for gas collisions at the end of each timestep.
Gas collisions rotate the trajectory to an uniformly distributed random
angle, but do not change the magnitude of the velocity. The rate of
gas collisions is determined by an exponential distribution with characteristic
timescale $t_{c}$. The walls of the 2D rectangular cell are Lambert
scattering with the scattering angle at each wall collision calculated
by $\theta=2\arcsin(\sqrt{Rand()})-\pi/2$, where $\theta$ is the
scattering angle relative to the normal of the wall and ranges from
$-\pi/2$ to $\pi/2$, and $Rand()$ is a function that returns a
uniformly distributed random number between 0 and 1, thereby giving
a cosine distributed $\theta$. Wall collision times are determined
by checking if between the i and the i+1 timestep the trajectory will
intersect with the wall, and if so calculating a wall collision at
the i\textsuperscript{th} time. Small timesteps relative to the ballistic
time are needed to accurately calculate the trajectories in the ballistic
limit. The simulated trajectories are used to calculate the position-position
correlation function. The spectrum is then determined by taking the
fast Fourier transform. The simulation only considers $\tau>0$, where
the full spectrum includes negative tau, therefore the real part of
the simulation is equal to 1/2 of $S_{rr}$. Like in our model, times
are measured in units of $t_{b}$ and the speed of the particles is
fixed to 1. Additionally the center of the cell is located at the
origin with walls at $\pm L/2$. 

Fitting our model to a 4000 particle simulation over 20000 0.0125$t_{b}$
timesteps of a 2x2 square cell in the ballistic limit ($\lambda=0.02$)
with Lambert scattering walls as shown in Fig. 4 we determine $\sigma_{square}=0.393$
and $L_{modsquare}=2.13$. When fitting our model we only use the
$n=\pm1$ terms, assuming the contribution from $\left|n\right|>1$
terms is negligible. To find the value of $\sigma_{v,a=\infty}$ that
best fits the simulation, our model (using $\sigma_{square}=0.393$
and $L_{modsquare}=2.13$) is fit to a 16000 particle 20000 timestep
ballistic limit simulation for a 2x8 rectangular cell ($a=4$) as
shown in Fig. 5. From this fit we find $\sigma_{v,a=\infty}=0.54$.
In all plots comparing the model and simulation only 1/2 of the $S_{rr}$
calculated from the model is plotted, and only $n=\pm1$ terms of
the calculation are used.

\begin{figure}[H]
\includegraphics[width=8.6cm]{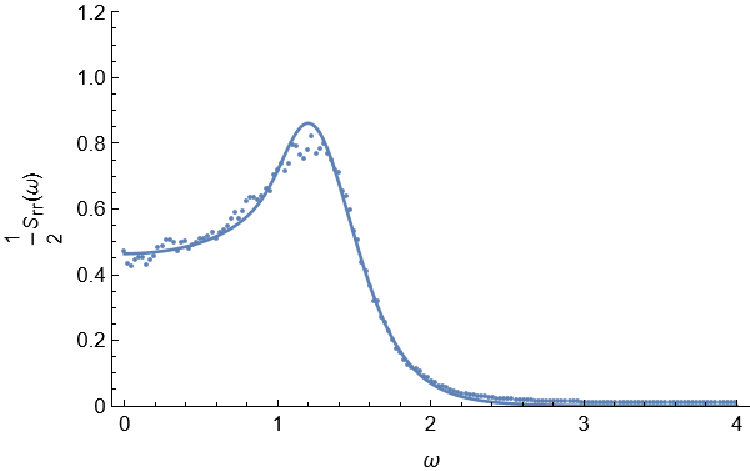}\caption{Simulation of the spectrum of the position-position auto-correlation
function for a 2x2 square with Lambert scattering walls in the ballistic
limit ($\lambda=0.02$) along with a a fit to the data using the model
described by Eq. (19) and Eqs. (30)-(34). From this we find the best
fit values for $\sigma_{square}$ and $L_{modsquare}$ are 0.393 and
2.13 respectively.}
 
\end{figure}

\begin{figure}
\includegraphics[width=8.6cm]{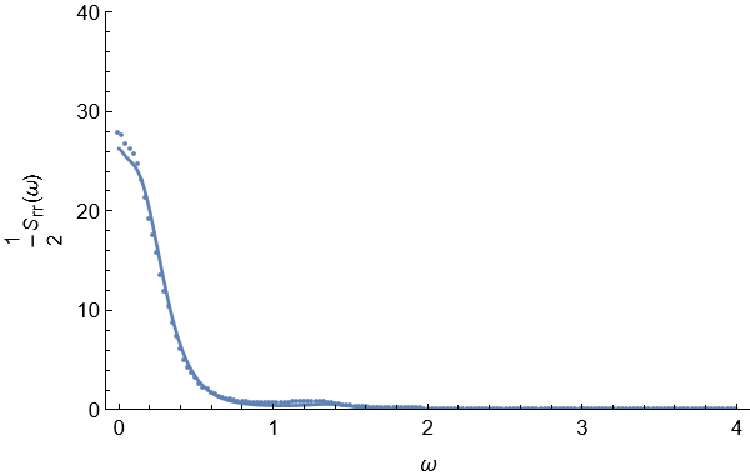}

\caption{Simulation of the spectrum of the position-position auto-correlation
function for a 2x8 rectangle with Lambert scattering walls in the
ballistic limit ($\lambda=0.02$) along with a a fit to the data using
the model described by Eq. (19) and Eqs. (30)-(34). From this we find
the best fit value for $\sigma_{v,a=\infty}=0.54$, given $\sigma_{square}=0.393$
and $L_{modsquare}=2.13$.}
\end{figure}

Using this model and the parameter values found from out fitting the
$a=1$ and $a=4$ cells, we find good agreement in the ballistic limit
for the entire range of aspect ratios from $a=1$ to $a=6$ as shown
in Fig. 6. 

\begin{figure}
\includegraphics[width=8.6cm]{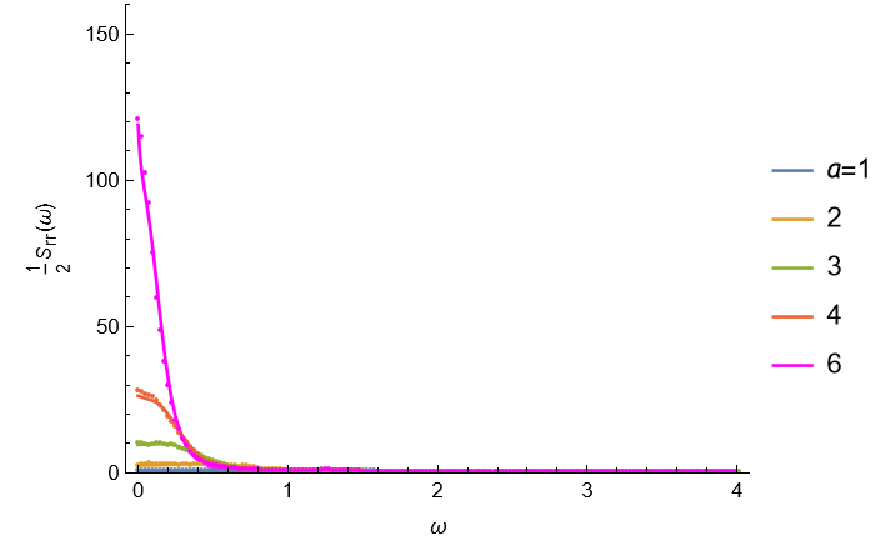}\caption{Simulations of the spectrum of the position-position auto-correlation
function between $a=1$ and $a=6$ in the ballistic limit ($\lambda=0.02$)
along with the model described in by Eq. (19) and Eqs. (30)-(34) for
the same aspect ratios. Model curves are plotted using $\sigma_{v,a=\infty}=0.54$,
$\sigma_{square}=0.393$, and $L_{modsquare}=2.13$.}
\end{figure}

Similarly we see good agreement going from the ballistic to diffusive
limit for all aspect ratios, especially for the case of the square
cell, with the largest disagreement between model and simulation occurring
around the $S_{xx}$ peak at higher aspect ratios. Fig. 7 shows comparisons
of our model to simulation from the ballistic to diffusive limit for
$a=1$ through 4. Including the $\left|n\right|>1$ terms in the initial
fitting may result in slightly different values of the fitted parameters.
For rectangular cells where the spectrum is dominated by $S_{yy}$
this change should be most noticeable in $\sigma_{v,a=\infty}$. Additionally
since we only consider $\left|n\right|=1$ the possibility of the
fitted parameters having $n$ dependence is not explored. 

\begin{figure*}
\subfloat[]{\includegraphics[width=8.6cm]{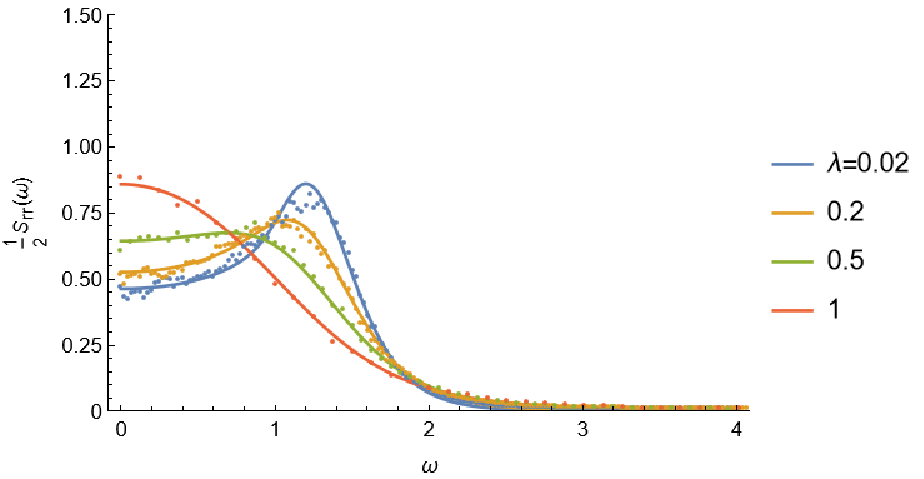}\includegraphics[width=8.6cm]{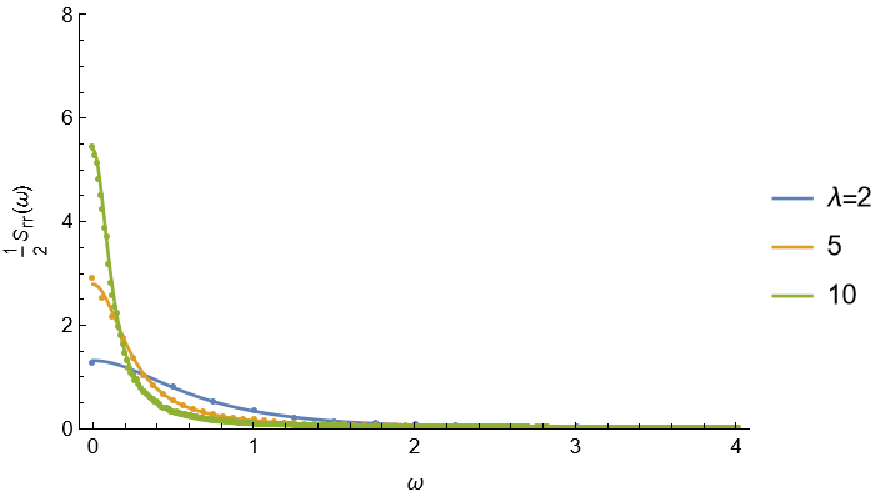}}

\subfloat[]{\includegraphics[width=8.6cm]{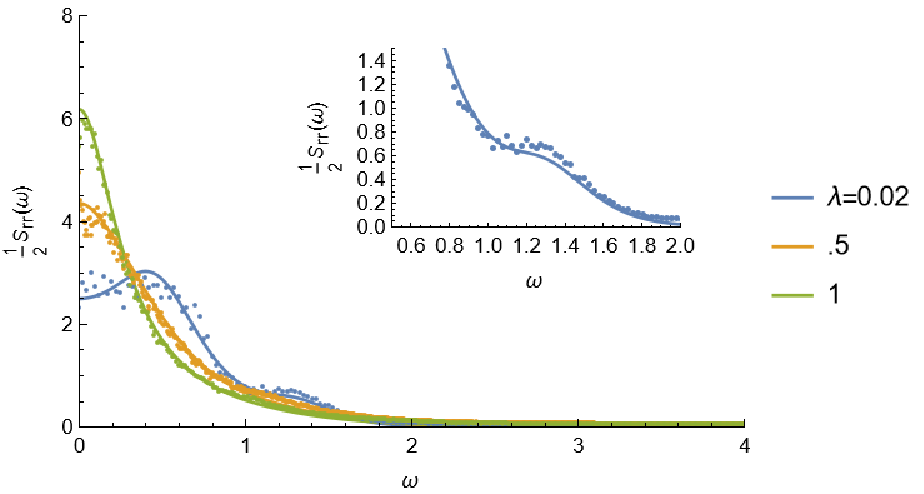}\includegraphics[width=8.6cm]{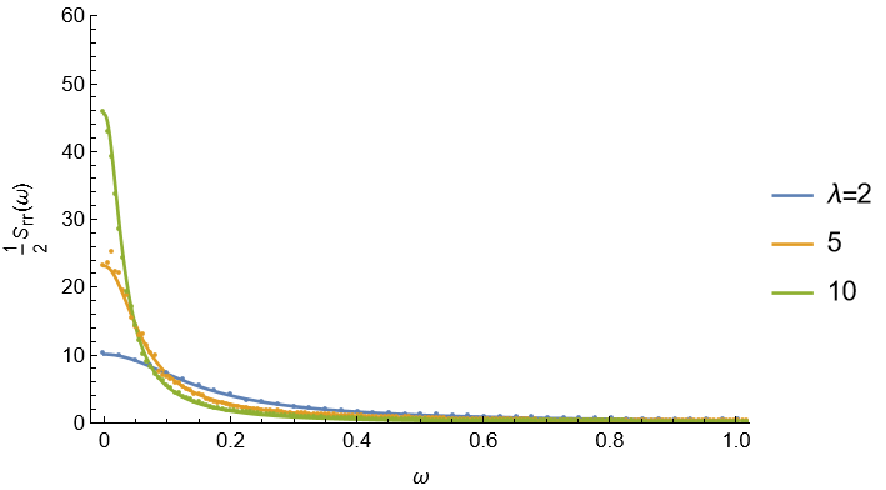}}

\subfloat[]{\includegraphics[width=8.6cm]{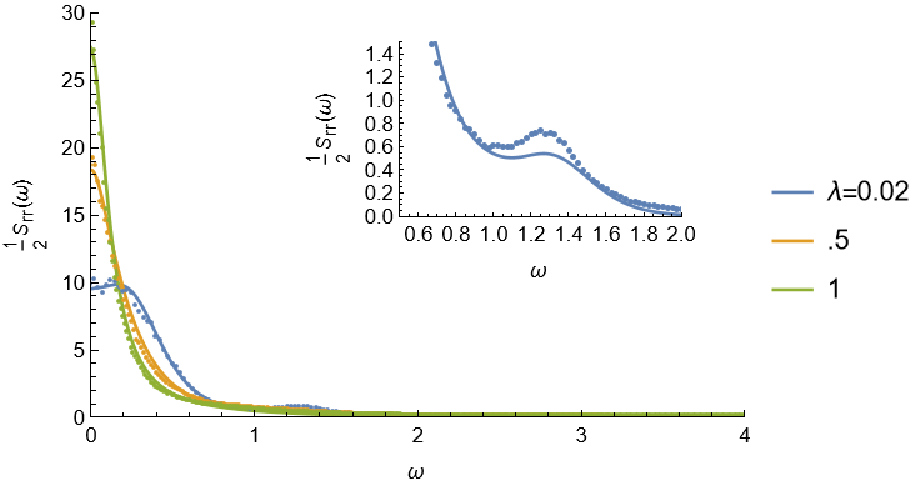}\includegraphics[width=8.6cm]{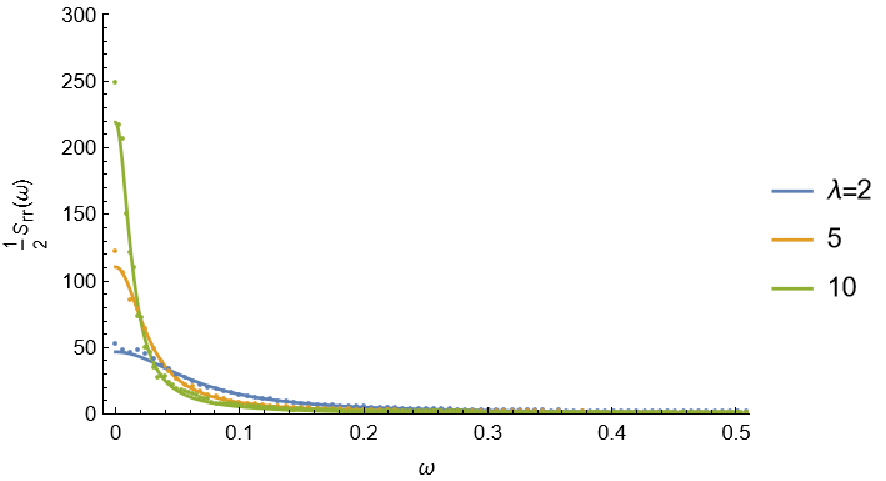}}

\subfloat[]{\includegraphics[width=8.6cm]{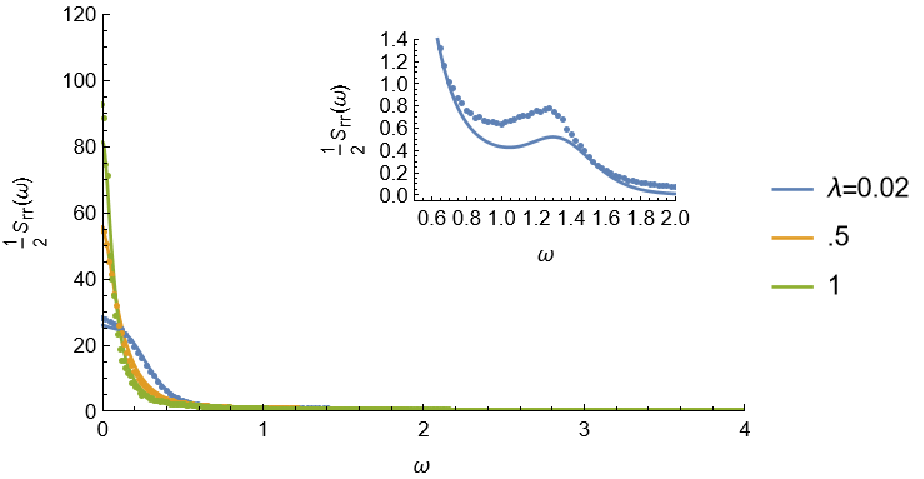}\includegraphics[width=8.6cm]{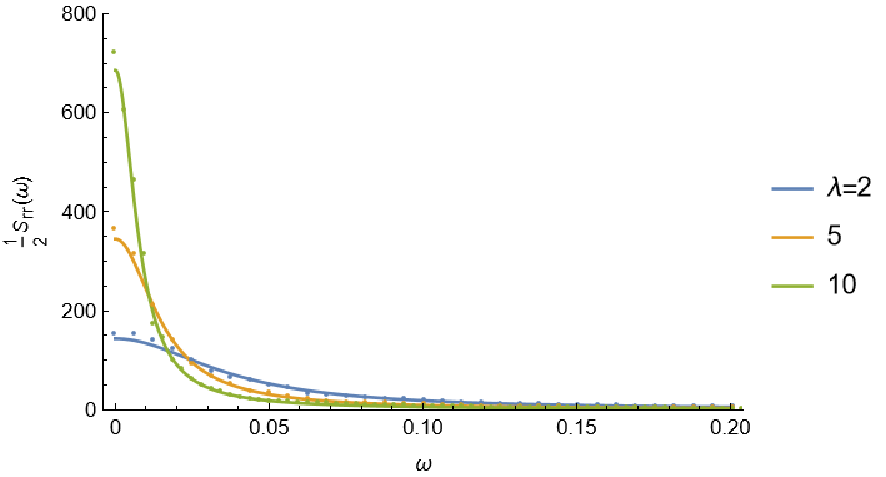}}\caption{Simulations of the spectrum of the position-position auto-correlation
function from the ballistic to diffusive regime along with the model
described in by Eq. (19) and Eqs. (30)-(34) . Model curves are plotted
using $\sigma_{v,a=\infty}=0.54$, $\sigma_{square}=0.393$, and $L_{modsquare}=2.13$.
Curves for aspect ratios of 1, 2, 3, and 4 are plotted in subfigures
(a), (b), (c), and (d) respectively, with $\lambda\protect\leq1$
on the left and $\lambda>1$ on the right. For the $\lambda\protect\leq1$
curves of subfigures (b),(c), and (d) we additionally plot a zoomed
in view of of the $\lambda=0.02$ curve to show the peak resulting
from $S_{xx}$ contribution to the correlation function. This peak
is suppressed at larger $\lambda$.}
\end{figure*}

\clearpage{}

\section*{Conclusion}

In this work we have expanded on the results of \citep{key-14,key-15}
and shown how the method of images can be used to model Lambert scattering
walls for rectangular cells with reasonable accuracy from the ballistic
to diffusive regime for aspect ratios between 1 and 6. This is done
by replacing the periodic extension in the method of images used in
\citep{key-14,key-15} by a Gaussian distribution of extensions characterized
by a standard deviation given by $\sigma_{square}$ and an average
$L_{mod}$. For rectangular cells a Gaussian spread in the speed is
also needed in the $S_{yy}$ spectrum characterized by standard deviation
$\sigma_{v}$ in order to account for the effects of an asymmetric
cell on the x and y components of the velocity after a wall collision.
It also should be noted that the Gaussian distributions used in this
model are only approximations. The true distributions shouldn't allow
negative values of $l$ or $v$. The resulting correlation function
is given by Eq. (19) and Eqs. (30)-(34). Comparing to simulation we
find $\sigma_{square}=0.393$, $L_{modx}=2.13$, and $\sigma_{v,a=\infty}=0.54$.
With only three free parameters, two of which describe the wall scattering
in a square cell, and a third that describes asymmetry in the mixing
of x and y components of the velocity due to wall collisions, this
models allows us to describe a remarkably wide range of aspect ratios
and gas collision rates, and reproduces the feature of two peaks in
the ballistic limit spectrum seen in simulations of rectangular cells.
The larger peak coming from the $S_{yy}$ contribution to the spectrum,
and the much smaller peak between $\omega=1.2$ and 1.4 coming from
$S_{xx}$. Although further work is needed to understand the differences
between model and simulation in the $S_{xx}$ peak shape. Given that
this model works by finding the distribution of triangle periodic
extensions that reproduces Lambert scattering, it may be possible
to reproduce other wall scattering conditions by finding the appropriate
values of $\sigma_{square}$, $L_{modx}$, and $\sigma_{v,a=\infty}$.
As such this work has applications for transport in mesoscopic systems
with with a variety of boundary conditions ranging from purely specular
to the purely Lambert scattering case discussed in this work. Additionally
the trajectory correlation functions in this work have applications
to the problem of spin relaxation and frequency shifts for polarized
gasses in inhomogeneous magnetic fields. As well as calculations of
the motional field ``false nEDM'' frequency shift seen in nEDM experiments.

\end{document}